\documentclass[draft]{agujournal2019}
\usepackage{url} 
\usepackage{lineno} 
\usepackage{soul} 
\usepackage{bm}  
\usepackage{gensymb}  
\usepackage{microtype}  
\usepackage{cleveref}  
\usepackage{xcolor}

\draftfalse
\journalname{Geophysical Research Letters}

\graphicspath{{./}}

\begin{document}

%
%


\title{Ensemble Covariance Propagation with Machine Learning Weather Prediction Models in a Realistic Data Assimilation System}

%
%




\authors{S. Toedtli\affil{1}, D. Ahijevych\affil{1}, W. Wang\affil{1}, and C. Snyder\affil{1}}


\affiliation{1}{National Science Foundation National Center for Atmospheric Research, Boulder, Colorado}




\correspondingauthor{Simon Toedtli}{stoedtli@ucar.edu}



\begin{keypoints}
\item A two-way coupled data assimilation system provides a clean test of covariance propagation with the machine learning model GraphCast
\item GraphCast matches a physics-based system in ensemble spread and covariance structure over a month of cycling with real observations
\item The GraphCast system matches or outperforms a physics-based system except near the model top and is a proof of concept for operational use
\end{keypoints}

%
%

%
%


\begin{abstract}  
Integrating machine learning weather prediction (MLWP) models into classical data assimilation (DA) workflows requires that they accurately evolve ensemble covariances.
We evaluate this capability in a two-way coupled DA system, where an ensemble variational (EnVar) component evolves the mean and an ensemble Kalman filter (EnKF) evolves the covariances.
The system is cycled over one month and assimilates a wide range of real-world observations, including microwave radiances.
Using the MLWP model GraphCast for the EnKF and the physics-based MPAS model for the EnVar yields ensemble spread and covariances comparable to a reference configuration in which MPAS evolves both components.
The resulting EnVar mean achieves consistently lower RMSE against GFS analyses, except near the model top, where spurious correlations degrade performance.
These results demonstrate that MLWP models can propagate covariance information in a realistic DA cycling system and provide a proof of concept for their integration into future operational ensemble systems.
\end{abstract}

\section*{Plain Language Summary}  
\noindent Weather forecasts begin from a best estimate of the current atmosphere, which is obtained by combining a short forecast with observations.
Forecast centers repeat this procedure every few hours with many slightly different forecasts, whose differences quantify the forecast uncertainty and how much each observation should correct its surroundings.
Running these forecasts with a traditional physics-based model is costly.
Machine learning (ML) weather models are much cheaper, but earlier attempts to use them in this process gave mixed results, and some forecasts became unstable after a few iterations.
A suspected cause is that ML models do not correctly capture how forecast uncertainties evolve.
We tested this by letting a physics-based model produce the best estimate of the atmosphere, while the ML model, GraphCast, evolved only the forecast uncertainties.
We ran this system for one month with a wide range of real-world observations, including satellite data.
It performed as well as, or slightly better than, a system using only the physics-based model at most levels of the atmosphere, showing that ML models can correctly evolve forecast uncertainties.
Because this setup resembles the systems used at forecast centers, the results show a practical path to cheaper everyday forecasting with ML models.

%
%


%


%
%
%
%

\section{Introduction}
Machine learning-based weather prediction (MLWP) models have improved rapidly in recent years \cite <see, for example,>[]{pathak_fourcastnet_2022,bi_accurate_2023,bonev_spherical_2023,lam_learning_2023,kochkov_neural_2024,price_probabilistic_2025}.  The best current MLWP models now surpass traditional, physics-based numerical models in many aspects of global medium-range forecasting \cite{lang_aifs_2024} and, once trained, can be run at a fraction of the computational cost.

A natural question is then whether a skillful MLWP model can replace traditional numerical models in the data-assimilation (DA) workflows that provide initial conditions for medium-range forecasts.
Several studies, using both variational and ensemble-based DA schemes, have performed idealized experiments with synthetic observations simulated from reanalyses and shown that MLWP models can cycle stably with DA (i.e. repeatedly alternate forecast and assimilation steps, see text S1 in the supplementary material for details) for extended periods \cite{xiao_fengwu-4dvar_2024,adrian_data_2025,kotsuki_ensemble_2025}.
When assimilating real observations, cycling DA results with MLWP models have been mixed. \citeA{slivinski_assimilating_2025} show that an Ensemble Kalman Filter (EnKF) assimilating surface pressure observations generates accurate global analyses when forecasts come from a physics-based model, but diverges when MLWP forecasts are used instead.
On the other hand, \citeA{takeshima_bridging_2025} and \citeA{huang_ensemble-based_2026} assimilate the global network of conventional observations and show that EnKFs with MLWP forecasts cycle stably, though they do not compare to the same DA system with forecasts from a physics-based model.

Because modern DA schemes rely on the forecast model to provide covariance information, one requirement for  MLWP models used in DA is that they should accurately capture the evolution of perturbations, such as analysis increments or ensemble deviations.
It is an open question whether MLWP models do this.  They clearly can degrade cross-variable relationships such as geostrophy \cite{bonavita_limitations_2024} and there is broad consensus that MLWP models trained with a root-mean-square error (RMSE) loss overly damp energy at small scales, with implications for perturbation growth \cite{lam_learning_2023,bonavita_limitations_2024,subich_fixing_2025}. In addition, gradients of forecast quantities with respect to initial conditions, which are enabled by the autodifferentiability of MLWP models, can exhibit unphysical features \cite{tian_evaluating_2026}, though other case studies find that MLWP gradients agree well with those from the adjoint of a physics-based numerical model \cite{bano-medina_are_2025}.
Moreover, and with most direct relevance to the present paper, \citeA{slivinski_assimilating_2025}, \citeA{kotsuki_ensemble_2025} and \citeA{takeshima_bridging_2025} all conclude that differences in their experiments between DA with MLWP forecasts or with physics-based numerical forecasts, including the divergence of cycling DA with MLWP models in some cases, stem from flaws in the way MLWP models evolve small perturbations.

This paper explores further the use of MLWP models in existing DA workflows. 
We begin from a cycling DA system that is credible, if not of operational quality, and that uses a state-of-the-art physics-based numerical model.  We configure the system, as is current practice at most operational centers for global weather prediction, to employ an ensemble-variational (EnVar) DA scheme, which minimizes a cost function that includes covariances estimated from a forecast ensemble, together with an EnKF, which produces initial conditions for the forecast ensemble.  This configuration allows the model used for ensemble forecasts to be different from that used for the ``deterministic” EnVar cycle. We test replacing the physics-based model with a MLWP model for the ensemble forecasts, thus cleanly isolating the effects of the MLWP model to the propagation of covariance information.  Our experiments also improve on previous work by using a more comprehensive suite of observations that includes microwave radiances from polar-orbiting instruments, which are the backbone of the global observing network.
We demonstrate that using an MLWP model to evolve the ensemble covariances can perform as well as, or better than, using a physics-based model.

The details of the forecast models, the DA system, and the assimilated observations are described in \cref{sec:methods}.
\Cref{sec:results} compares the performance of the physics-based and MLWP-based DA systems and characterizes the evolution of the deterministic background quality (\cref{sec:verification}), RMS-spread statistics (\cref{sec:spread}) and single-observation experiments (\cref{sec:singleObs}).
Broader implications are discussed in \cref{sec:conclusions} and additional details and statistics are provided in the supplementary material.

\section{Methods}  \label{sec:methods}

\subsection{Forecast models}
The two forecast models considered in this study are MPAS \cite{skamarock_multiscale_2012} and GraphCast \cite<referred to as GC hereafter,>[]{lam_learning_2023}.
MPAS is a physics-based model, discretized on an unstructured horizontal mesh with a hybrid terrain-following vertical coordinate, that has been used previously in cycling DA experiments \cite<>[among others]{liu_data_2022,sun_all-sky_2025}.
GC, by contrast, is an MLWP model whose state variables are defined on a structured latitude–longitude mesh and a discrete set of vertical pressure levels.
Since the pretrained GC model has a fixed mesh, MPAS is configured to match GC's model top and horizontal resolution as closely as possible.

Specifically, the MPAS configuration adopted here uses a global domain with a quasi-uniform $120$ km horizontal mesh, a $22.6$ km model top with a gravity wave absorbing layer starting at $18.5$ km, and $47$ hybrid vertical levels.
This model top is lower than in previous studies, and chosen to approximately match the $50$ hPa top of GC.
For GC, the small configuration with $1\degree$ horizontal resolution and $13$ vertical pressure levels is used.
The two models have comparable horizontal resolution at the equator, but GC has a finer longitudinal grid spacing toward the poles.

\subsection{Cycling workflow}
\begin{figure}
    \centering
    \includegraphics[width=\textwidth]{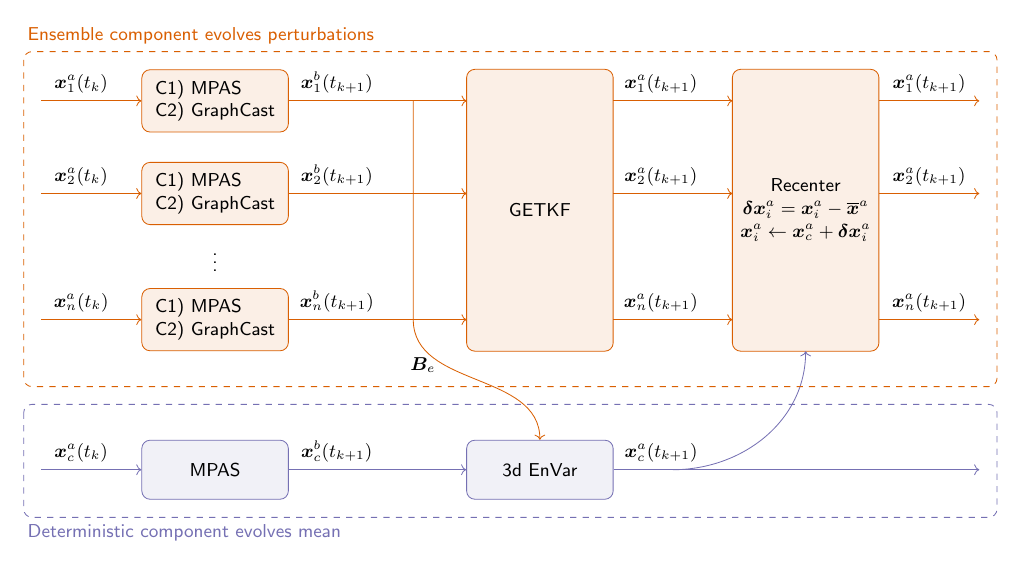}
    \caption{Schematic of the data assimilation cycling workflow for the two-way coupled ensemble (top, orange) and deterministic (bottom, purple) system. Arrows indicate the flow of information, with the following notation: background $\bm{x}^b$, analysis $\bm{x}^a$, perturbation $\bm{\delta x}$, mean $\overline{\bm{x}}$, and ensemble background error covariance $\bm{B}_e$. The subscript $i = 1, \ldots, n$ indexes ensemble members, and the subscript $c$ denotes the deterministic control state. Configurations C1 and C2 differ in the forecast model for the ensemble component.}
    \label{fig:daWorkflow}
\end{figure}

The DA cycling workflow consists of concurrently running, two-way coupled EnKF and EnVar systems, as illustrated in \cref{fig:daWorkflow}.
The EnVar uses the EnKF background ensemble to estimate the background error covariance, while the EnKF analysis ensemble is recentered on the EnVar analysis at each DA cycle.
This recentering is central to the design of this study, since it restricts the role of the ensemble forecast model to evolving ensemble perturbations around the mean of the EnVar component.
The EnKF system is formulated in gain form (GETKF) with model-space localization in the vertical coordinate \cite{bishop_gain_2017} and consists of $20$ ensemble members.
Further details of both DA components are provided in the supplementary material (text S2).

The forecast model for each DA component can be chosen independently, and two configurations are considered for this study, as indicated in \cref{fig:daWorkflow}.
Configuration C1 uses the physics-based MPAS model for both the EnKF and EnVar forecasts and serves as a physics-based benchmark. 
Configuration C2 replaces MPAS with GC for the EnKF forecast and retains the physics-based MPAS model for the EnVar forecast.
A comparison between configurations C1 and C2 provides a direct test of whether GC can accurately evolve ensemble covariances, because the two configurations differ only in the model used to propagate the ensemble.

The ensemble inflation for each configuration is tuned independently (see text S3 for details), yet the optimal inflation parameters are identical for configurations C1 and C2.
This contrasts with previous studies, where MLWP-based ensembles require additional inflation to cycle stably \cite{kotsuki_ensemble_2025,huang_ensemble-based_2026}.

All data assimilation steps are performed in MPAS-JEDI on the native MPAS mesh, regardless of the choice of ensemble forecast model.
In configuration C2, this requires interpolating between the MPAS and GC meshes before and after each ensemble forecast step, and supplementing the GC state with the soil variables it lacks.
Details of this procedure are provided in text S4.

\subsection{Data assimilation experiments}
Configurations C1 and C2 are cycled every six hours over a one-month period (2018-04-15T00Z – 2018-05-14T18Z).
The initial background at 2018-04-15T00Z is obtained from a 6 h MPAS forecast, initialized from GFS (EnVar component) and GEFS (EnKF), respectively.
Subsequent DA cycles generate $6$ h forecasts with the designated forecast model, initialized from the latest analysis (MPAS) or the two most recent analyses (GC, which requires two input states).

Each DA component assimilates the same set of observations, which represents a substantial subset of the global observation network, including conventional observations, satellite-derived atmospheric motion vectors, GNSS-RO bending angle profiles, and microwave radiances (see text S5 and table S1 for details).
To the authors' knowledge, this marks the first assimilation of microwave radiances into an MLWP-based forecast via a traditional radiative transfer model and DA framework, rather than an end-to-end ML-based assimilation approach that maps observations directly to an analysis \cite<as used, for example, in>[]{xu_fuxi-da_2025,sun_data--forecast_2025}.
Observations are ingested in raw form and quality control, observation error modeling, and radiance bias correction are all performed within MPAS-JEDI.

\section{Results}  \label{sec:results}

\subsection{Deterministic component: model-space verification}  \label{sec:verification}
Because the EnVar system relies on the EnKF ensemble to estimate the background error covariance at each analysis step (see \cref{fig:daWorkflow}), a verification of the $6$ h EnVar background provides a global assessment of the EnKF ensemble.
This section presents this background verification against the GFS analysis, which is taken as the ground truth.

\begin{figure}
    \centering
    \includegraphics[width=0.8\textwidth]{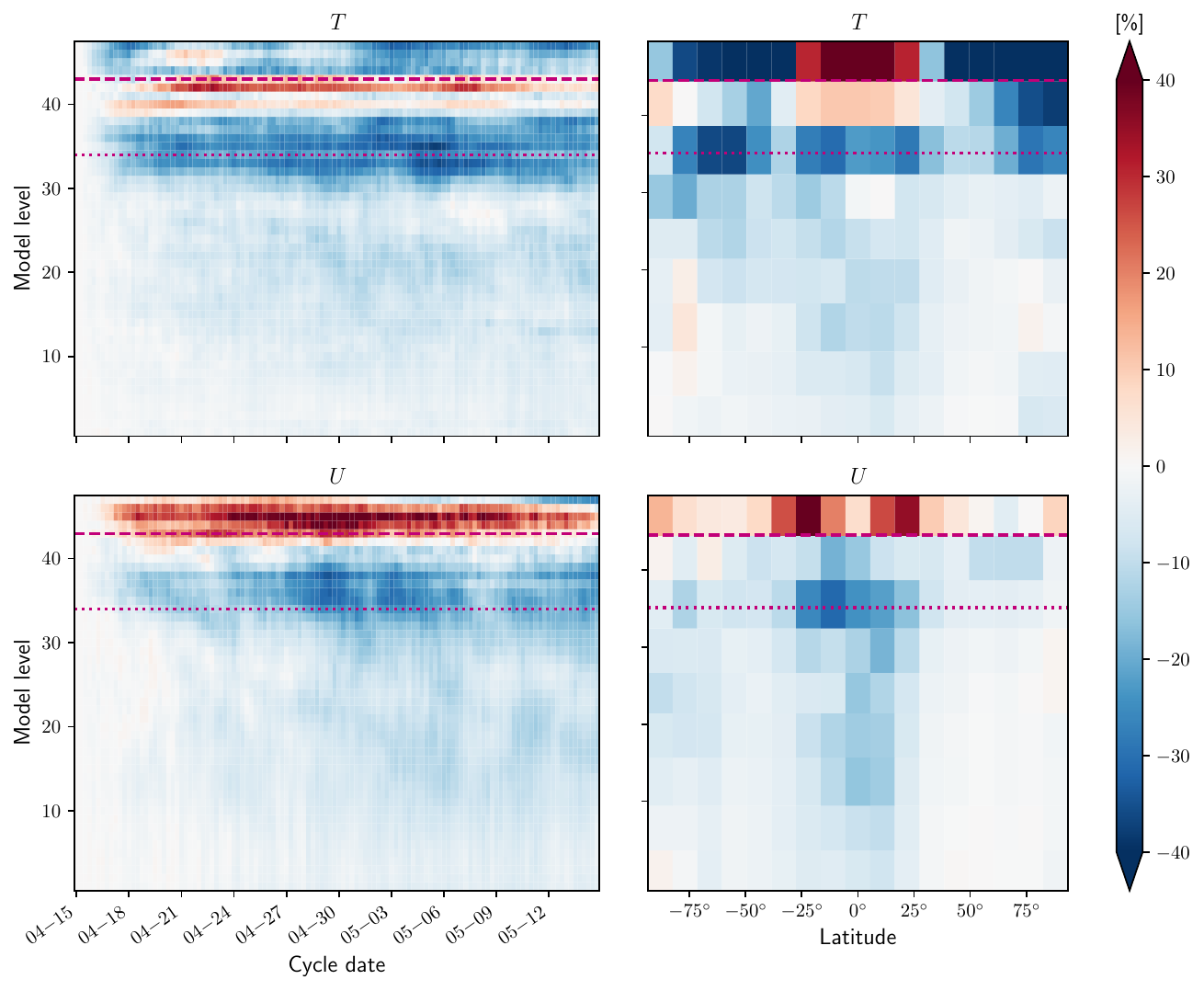}
    \caption{Verification of the $6$ h EnVar background against the GFS analysis, shown as the relative RMSE difference between configurations C2 and C1, $\mathrm{rms}(\mathrm{C2}) / \mathrm{rms}(\mathrm{C1}) - 1$. Negative values (blue) indicate improvement of C2 over C1. Top row: temperature ($T$), bottom row: zonal wind ($U$). Left column: RMSE averaged horizontally, right column: RMSE averaged over longitude and time. The dashed and dotted horizontal lines indicate, respectively, the approximate location of the GC model top and the highest level in the EnVar analysis that can be directly influenced by observations above the GC model top.}
    \label{fig:backgroundVerif}
\end{figure}

To simplify the comparison between the two configurations, \cref{fig:backgroundVerif} shows the model-space verification in terms of the relative RMSE difference, $\mathrm{rms}(\mathrm{C2}) / \mathrm{rms}(\mathrm{C1}) - 1$, so that negative values (blue) indicate better performance of C2. 
The comparison reveals two prominent differences between the two configurations.
First, below model level $40$, configuration C2 exhibits smaller RMSEs than C1.
This improvement persists over the entire month of cycling and grows slowly with time (left column).
When binned by model level and latitude (right column), the RMSE reduction is apparent at almost all latitudes and model levels, with the largest improvements over the tropics and southern hemisphere, and smaller improvements over the northern hemisphere.
Below model level 40, the MLWP-based system is therefore at least as accurate as the physics-based baseline.

The second, contrasting observation is a pronounced deterioration in configuration C2 that develops above level 40 after approximately two days of cycling and persists throughout the month.
Three factors likely contribute.
First, the mismatch between the MPAS and GC model tops:
the physical height of GC’s 50 hPa top varies with location and time and can be as low as 19 km.
When GC forecasts are interpolated to the MPAS mesh, all quantities above the 50 hPa isobar must be extrapolated, which introduces additional error in the ensemble fields compared to the interpolated levels below.
The dashed magenta line in \cref{fig:backgroundVerif} marks the lowest MPAS level at which this can occur. Extrapolation errors can also affect the DA at lower levels through vertical ensemble covariances, but only within the 6 km vertical localization distance used in both the EnVar and GETKF systems (dotted line).
Second, MPAS applies a gravity wave damping layer that damps vertical motions above 18.5 km (approximately the dashed magenta line), whereas GC has no analogous damping, so the two systems represent different dynamics there.
Third, spurious sensitivities near the model top, which have been reported for various MLWP models in both the linear and nonlinear regimes, may also contribute \cite{bano-medina_are_2025,tian_evaluating_2026}.

We have not isolated the root cause, and a combination of these factors likely contributes to the observed deterioration.
Even so, away from the immediate vicinity of the GC model top, the MLWP ensemble performs at least as well as, and in general better than, the MPAS ensemble.

\subsection{Ensemble component: model-space RMS-spread statistics}  \label{sec:spread}
\begin{figure}
    \centering
    \includegraphics[width=0.7\linewidth]{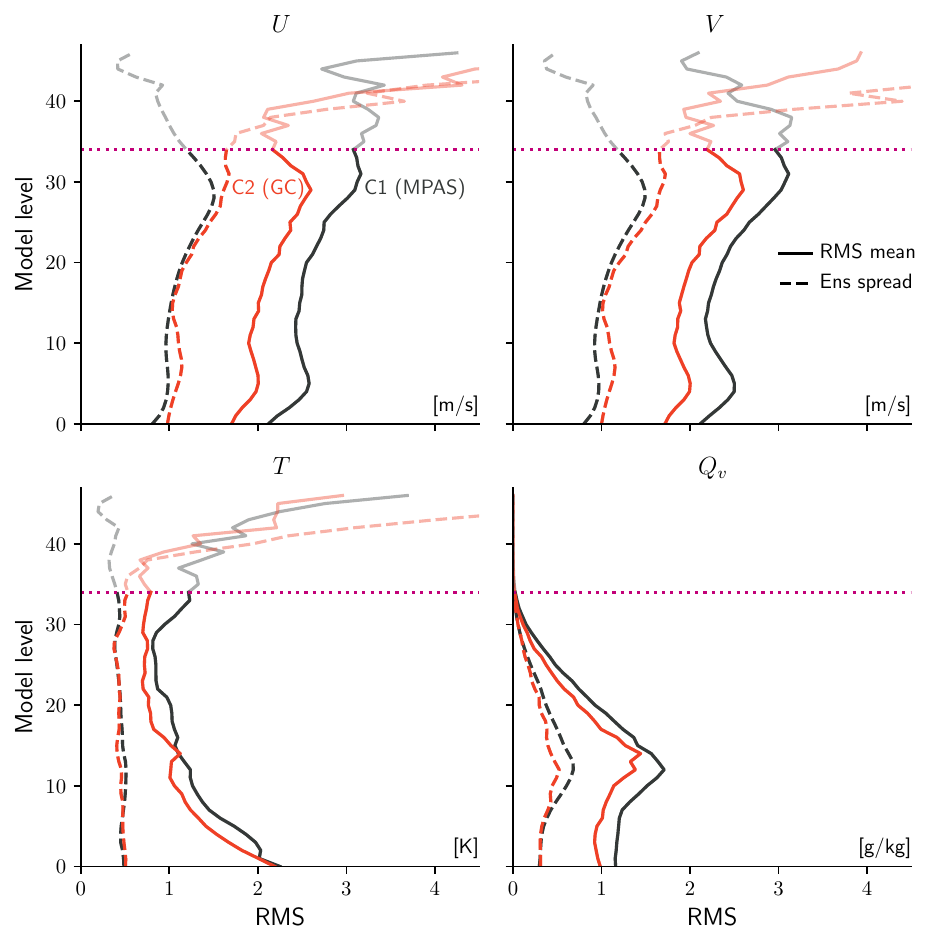}
    \caption{RMS-spread statistics for the EnKF background ensemble. Solid lines show the RMSE of the ensemble mean against the GFS analysis, dashed lines represent the ensemble spread. Black: configuration C1 (MPAS), red: configuration C2 (GC). Results are averaged horizontally and over the last week of the cycling period. Model levels above the dotted horizontal line are potentially affected by spurious correlations near the GC model top.}
    \label{fig:ensSpread}
\end{figure}

Ensemble spread diagnostics provide a second, complementary assessment of the MLWP model's ability to propagate covariances.
\Cref{fig:ensSpread} compares the spread of the background ensembles against the RMSE of the ensemble mean in model space, with the GFS analysis taken as ground truth.
Model levels above the dotted line are potentially affected by extrapolation above the GC model top (see previous section) and are therefore shown muted.

Below the dotted horizontal line, the two ensembles have very similar spread (dashed lines) across all variables and levels.
This has two important implications.
First, it demonstrates that GC can maintain an ensemble spread comparable to a physics-based model over many DA cycles.
Second, it rules out the possibility that the RMSE differences seen in \cref{fig:backgroundVerif} below this level are driven by systematic differences in how the two systems weight the background against observations.
This suggests instead that the lower RMSE of the GC configuration reflects an improved ensemble covariance.

Figures S1 and S2 further show that the two ensemble spreads remain approximately constant over time and have a similar spatial structure, even if local differences exist.
For example, configuration C2 typically shows larger spread in the polar regions and lacks some of the small-scale features seen in C1 over the mid-latitudes and tropics.
This suggests that the two configurations may weigh the background and observations somewhat differently at specific locations, but they assign comparable weight to each on average.

Comparing the ensemble spread to the RMSE of the mean (solid lines) shows that both systems are under-dispersive.
This likely arises from the known bias of the EnKF update, which underestimates the analysis variance, and from unaccounted model error in the forecast step, which are not specific to either forecast model.
Under-dispersive ensembles are common in ensemble DA and have been reported in previous MPAS-JEDI systems (see e.g. fig. 1 in \citeA{guerrette_data_2023} and fig. 2 in \citeA{sun_all-sky_2025}).
Nevertheless, the gap between spread and RMSE is generally smaller for configuration C2 (GC), owing to the higher-quality ensemble mean discussed in the previous section.

\subsection{Ensemble component: covariance structure}  \label{sec:singleObs}
As a final, qualitative assessment of the ensemble background error covariance, this section examines its structure through single-observation experiments conducted at the end of the cycling period (2018-05-15T00Z).
Results are shown for single temperature observation with a 1 K innovation and a 1 K observation-error standard deviation.
We tested multiple observation locations and present here representative results for an observation placed over the North Pacific ($40\degree$N, $175\degree$W) at approximately $702$ hPa (model level 17).

\begin{figure}
    \centering
    \includegraphics[width=0.95\linewidth]{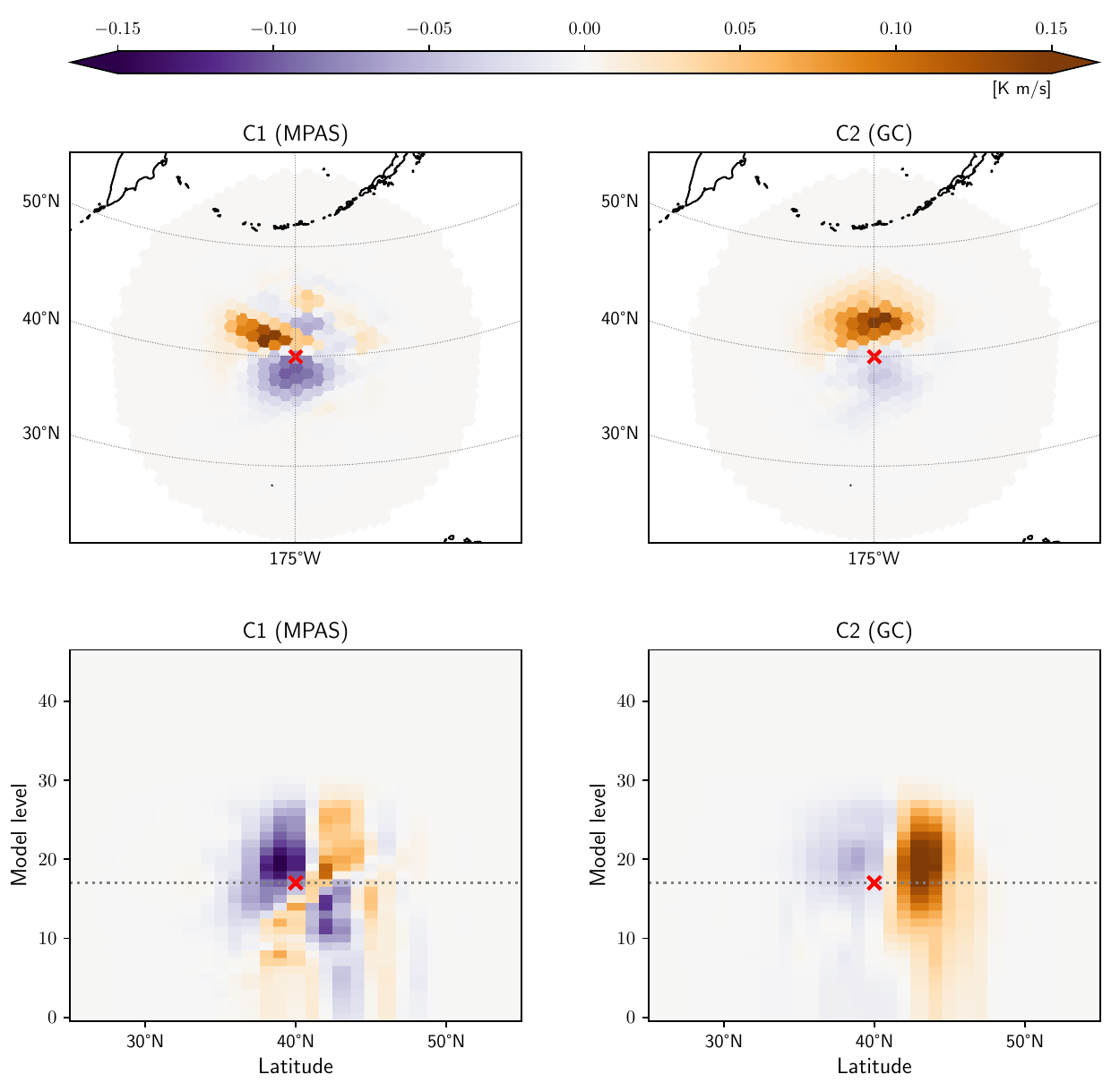}
    \caption{Single-observation experiment at the end of the cycling  (2018-05-15T00): cross-covariance between zonal wind and temperature, $\mathrm{cov}(U(\bm{r}), T(\bm{r}_M))$, where $\bm{r}_M$ denotes the measurement location. The measurement location is at $40\degree$ N, $175\degree$ W, $702$ hPa and indicated by the $\boldsymbol{\times}$ marker. Left column: C1 (MPAS); right column: C2 (GC). Top row: horizontal cross section at the measurement height; bottom row: vertical cross section along $175\degree$ W.}
    \label{fig:covUT}
\end{figure}

\Cref{fig:covUT} shows the cross-covariance between zonal wind and the temperature at the measurement location.
Both configurations show an anti-cyclonic wind pattern at and above the measurement location.
C1 additionally shows a weak cyclonic pattern below the measurement location, which is absent in C2.
This anti-cyclonic and cyclonic pattern is consistent with the thermal-wind relation in the mid-latitudes, and a similar structure, with location-specific variations, is found at other measurement locations as well (not shown).
Beyond this shared large-scale structure, the spatial arrangement, magnitude, and small-scale details of the two covariances differ.
These differences likely reflect a combination of genuine differences between the underlying ensembles and sampling uncertainty from the small ensemble size, and the comparison should therefore be interpreted qualitatively.

The autocovariance of the temperature field with the measurement shows a similar picture (Fig. S3): the spatial structure is again similar between the two configurations, but the magnitude in C2 is somewhat smaller than in C1.
This magnitude difference is location-dependent, as illustrated in the top row of Fig. S2, which shows the temperature variance, i.e. the autocovariance at zero separation, at the measurement level (model level 17) across the globe.

The single-observation experiments indicate that GC can maintain a realistic covariance structure over many DA cycles, consistent with the spread results discussed in the previous section.
There is no evidence of grid-scale noise build-up or weakening of the cross-covariances over time, in contrast to the failure mode reported in \citeA{slivinski_assimilating_2025}.
Although the covariances differ in magnitude, shape, and small-scale details, the lower background errors of configuration C2 (\cref{sec:verification}) suggest that the MLWP-based covariances propagate observation information at least as effectively as the physics-based ones.

\section{Conclusions}  \label{sec:conclusions}
Integration of MLWP models into classical DA workflows requires that they accurately evolve perturbations, such as analysis increments or ensemble member differences.
This study tested their capability to evolve ensemble perturbations that provide ensemble covariance information within a realistic, two-way coupled EnVar–EnKF DA system.
In this system, the EnVar component used the EnKF ensemble to estimate background error covariances, and the EnKF analysis ensemble was recentered on the EnVar analysis at each cycle.
A baseline configuration (C1) used a physics-based model (MPAS) to evolve both the EnVar and EnKF forecasts, while a comparison configuration (C2) used an MLWP model (GraphCast) to evolve the EnKF members but retained MPAS for the EnVar forecast.
This design deliberately limited the MLWP model's role to evolving ensemble perturbations and the covariance information they provide, so that a comparison of C1 and C2 provided a clean test of this capability.
Each DA component assimilated a substantial subset of the global observation network, including conventional observations, satellite-derived atmospheric motion vectors, GNSS-RO bending angle profiles, and microwave satellite radiances.

Configuration C2 cycled stably over an entire month and achieved a performance comparable to, or marginally better than, the physics-based benchmark C1 at most model levels.
The exception was near the model top, where artifacts from interpolating between the two model meshes degraded the quality of configuration C2.
This degradation could likely be reduced by more closely aligning the meshes of the physics-based and MLWP models, as done, for example, in AICON \cite{goecke_aicon_2026} and MAPCast \cite{wang_mapcast_2026}, or by employing the so-called hybrid EnVar available in JEDI \cite<see e.g.>[]{jung_three-dimensional_2024} and relying only on the parametric covariances at levels where the MLWP ensemble would need to be extrapolated.
Ensemble spread and single-observation experiments showed that the two systems produce comparable background error covariances on average, with local differences in covariance structure and magnitude.
These differences enable configuration C2 to better weigh and propagate observation information, which lead to the better background fit to GFS analyses.
Taken together, these results demonstrate that GraphCast can provide physically meaningful covariance information within a realistic DA cycling system.

A unique aspect of this study is the operationally relevant workflow, which separates the evolution of the ensemble mean and covariance.
Operational experience suggests that the quality of the background error covariances is often secondary to the quality of the mean forecast \cite<e.g.>[where the latter uses an even simpler background-covariance model such as 3DVar]{buehner_ensemble_2017,lorenc_computing_2017,parrish_national_1992} and the strong performance of the present configuration may partially reflect the benefit of limiting the MLWP model to the ensemble evolution.
A future configuration in which GraphCast also provides the EnVar forecast would offer a controlled test of the mean forecast's contribution to the overall performance.

The present study extends and contrasts with the existing literature with respect to DA system stability and performance, as well as ensemble inflation.
Like the MLWP-based filters of \citeA{kotsuki_ensemble_2025,takeshima_bridging_2025,huang_ensemble-based_2026} and the stabilized filter of \citeA{slivinski_assimilating_2025}, the present configuration cycles stably throughout the month-long experiment.
This provides further evidence that the diverging filters of \citeA{slivinski_assimilating_2025} based on MLWP model forecasts are the exception rather than the norm.
Compared to all of these previous studies, however, the present MLWP-based configuration did not require additional ensemble inflation and is the first to perform as well as, or better than, a physics-based benchmark.

Because the present configuration with its two-way coupled DA components and a wide variety of assimilated observations, including satellite radiances, closely mirrors an operational setup, these results constitute a proof of concept that MLWP models can supply covariance information within future operational ensemble systems as well.

\nocite{courtier_strategy_1994}
\nocite{derber_global_1989}
\nocite{gaspari_construction_1999}
\nocite{hunt_efficient_2007}
\nocite{whitaker_evaluating_2012}
\nocite{yessad_full-pos_2019}
\nocite{zhang_impacts_2004}

\section*{Open Research Section}
\noindent Observational data from the National Centers for Environmental Prediction (NCEP) used in the data assimilation experiments are available from the NSF NCAR Geoscience Data Exchange at \url{https://doi.org/10.5065/Z83F-N512}, \url{https://doi.org/10.5065/39C5-Z211}, and \url{https://doi.org/10.5065/DWYZ-Q852}.
The GFS analyses used for the EnVar initial conditions, model-space verification, and supplementing the GraphCast state are available from NOAA at \url{https://registry.opendata.aws/noaa-gfs-bdp-pds}, while the GEFS analyses used for the EnKF initial conditions are available at \url{https://registry.opendata.aws/noaa-gefs}.

All atmospheric models, the data assimilation framework, and the workflow engine used in this study are developed openly.
The data assimilation framework, the physics-based MPAS model, and the observation conversion tools are provided through MPAS-JEDI (\url{https://github.com/JCSDA/mpas-jedi}).
GraphCast is run through the Earth2Studio package (\url{https://github.com/NVIDIA/earth2studio}), and the data assimilation workflow is orchestrated with MPAS-Workflow (\url{https://github.com/NCAR/MPAS-Workflow}).

\section*{Conflict of Interest declaration}
\noindent The authors declare there are no conflicts of interest for this manuscript.

\acknowledgments
The authors thank BJ Jung and Craig Schwartz for helpful discussions.
This material is based upon work supported by the NSF National Center for Atmospheric Research (NCAR), which is a major facility sponsored by the U.S. National Science Foundation under Cooperative Agreement No. 1852977.
We acknowledge high-performance computing support from the Derecho system (doi:10.5065/qx9a-pg09) provided by NSF NCAR.
Additional funding was provided by the United States Air Force under Grant FA8730-21-C-0008.
During manuscript preparation, the authors used Claude Sonnet 5.5 (Anthropic) to improve the grammar, clarity, and flow of text written by the authors.

%
%

\bibliography{bibliography}

@article{skamarock_multiscale_2012,
	title = {A {Multiscale} {Nonhydrostatic} {Atmospheric} {Model} {Using} {Centroidal} {Voronoi} {Tesselations} and {C}-{Grid} {Staggering}},
	volume = {140},
	doi = {10.1175/MWR-D-11-00215.1},
	number = {9},
	journal = {Monthly Weather Review},
	publisher = {American Meteorological Society},
	author = {Skamarock, William C. and Klemp, Joseph B. and Duda, Michael G. and Fowler, Laura D. and Park, Sang-Hun and Ringler, Todd D.},
	year = {2012},
	note = {Place: Boston MA, USA},
	pages = {3090 -- 3105},
}

@article{liu_data_2022,
	title = {Data assimilation for the {Model} for {Prediction} {Across} {Scales} – {Atmosphere} with the {Joint} {Effort} for {Data} assimilation {Integration} ({JEDI}-{MPAS} 1.0.0): {EnVar} implementation and evaluation},
	volume = {15},
	copyright = {https://creativecommons.org/licenses/by/4.0/},
	issn = {1991-9603},
	shorttitle = {Data assimilation for the {Model} for {Prediction} {Across} {Scales} – {Atmosphere} with the {Joint} {Effort} for {Data} assimilation {Integration} ({JEDI}-{MPAS} 1.0.0)},
	doi = {10.5194/gmd-15-7859-2022},
	language = {en},
	number = {20},
	journal = {Geoscientific Model Development},
	author = {Liu, Zhiquan and Snyder, Chris and Guerrette, Jonathan J. and Jung, Byoung-Joo and Ban, Junmei and Vahl, Steven and Wu, Yali and Trémolet, Yannick and Auligné, Thomas and Ménétrier, Benjamin and Shlyaeva, Anna and Herbener, Stephen and Liu, Emily and Holdaway, Daniel and Johnson, Benjamin T.},
	month = oct,
	year = {2022},
	pages = {7859--7878},
}

@article{guerrette_data_2023,
	title = {Data assimilation for the {Model} for {Prediction} {Across} {Scales} – {Atmosphere} with the {Joint} {Effort} for {Data} assimilation {Integration} ({JEDI}-{MPAS} 2.0.0-beta): ensemble of {3D} ensemble-variational ({En}-{3DEnVar}) assimilations},
	volume = {16},
	copyright = {https://creativecommons.org/licenses/by/4.0/},
	issn = {1991-9603},
	shorttitle = {Data assimilation for the {Model} for {Prediction} {Across} {Scales} – {Atmosphere} with the {Joint} {Effort} for {Data} assimilation {Integration} ({JEDI}-{MPAS} 2.0.0-beta)},
	doi = {10.5194/gmd-16-7123-2023},
	language = {en},
	number = {23},
	journal = {Geoscientific Model Development},
	author = {Guerrette, Jonathan J. and Liu, Zhiquan and Snyder, Chris and Jung, Byoung-Joo and Schwartz, Craig S. and Ban, Junmei and Vahl, Steven and Wu, Yali and Baños, Ivette Hernández and Yu, Yonggang G. and Ha, Soyoung and Trémolet, Yannick and Auligné, Thomas and Gas, Clementine and Ménétrier, Benjamin and Shlyaeva, Anna and Miesch, Mark and Herbener, Stephen and Liu, Emily and Holdaway, Daniel and Johnson, Benjamin T.},
	month = dec,
	year = {2023},
	pages = {7123--7142},
}

@article{hunt_efficient_2007,
	title = {Efficient data assimilation for spatiotemporal chaos: {A} local ensemble transform {Kalman} filter},
	volume = {230},
	copyright = {https://www.elsevier.com/tdm/userlicense/1.0/},
	issn = {01672789},
	shorttitle = {Efficient data assimilation for spatiotemporal chaos},
	doi = {10.1016/j.physd.2006.11.008},
	language = {en},
	number = {1-2},
	journal = {Physica D: Nonlinear Phenomena},
	author = {Hunt, Brian R. and Kostelich, Eric J. and Szunyogh, Istvan},
	month = jun,
	year = {2007},
	pages = {112--126},
}

@article{lam_learning_2023,
	title = {Learning skillful medium-range global weather forecasting},
	volume = {382},
	issn = {0036-8075, 1095-9203},
	doi = {10.1126/science.adi2336},
	language = {en},
	number = {6677},
	journal = {Science},
	author = {Lam, Remi and Sanchez-Gonzalez, Alvaro and Willson, Matthew and Wirnsberger, Peter and Fortunato, Meire and Alet, Ferran and Ravuri, Suman and Ewalds, Timo and Eaton-Rosen, Zach and Hu, Weihua and Merose, Alexander and Hoyer, Stephan and Holland, George and Vinyals, Oriol and Stott, Jacklynn and Pritzel, Alexander and Mohamed, Shakir and Battaglia, Peter},
	month = dec,
	year = {2023},
	pages = {1416--1421},
}

@article{kochkov_neural_2024,
	title = {Neural general circulation models for weather and climate},
	volume = {632},
	issn = {0028-0836, 1476-4687},
	doi = {10.1038/s41586-024-07744-y},
	language = {en},
	number = {8027},
	journal = {Nature},
	author = {Kochkov, Dmitrii and Yuval, Janni and Langmore, Ian and Norgaard, Peter and Smith, Jamie and Mooers, Griffin and Klöwer, Milan and Lottes, James and Rasp, Stephan and Düben, Peter and Hatfield, Sam and Battaglia, Peter and Sanchez-Gonzalez, Alvaro and Willson, Matthew and Brenner, Michael P. and Hoyer, Stephan},
	month = aug,
	year = {2024},
	pages = {1060--1066},
}

@article{jung_three-dimensional_2024,
	title = {Three-dimensional variational assimilation with a multivariate background error covariance for the {Model} for {Prediction} {Across} {Scales} – {Atmosphere} with the {Joint} {Effort} for {Data} assimilation {Integration} ({JEDI}-{MPAS} 2.0.0-beta)},
	volume = {17},
	copyright = {https://creativecommons.org/licenses/by/4.0/},
	issn = {1991-9603},
	doi = {10.5194/gmd-17-3879-2024},
	language = {en},
	number = {9},
	journal = {Geoscientific Model Development},
	author = {Jung, Byoung-Joo and Ménétrier, Benjamin and Snyder, Chris and Liu, Zhiquan and Guerrette, Jonathan J. and Ban, Junmei and Baños, Ivette Hernández and Yu, Yonggang G. and Skamarock, William C.},
	month = may,
	year = {2024},
	pages = {3879--3895},
}

@article{gaspari_construction_1999,
	title = {Construction of correlation functions in two and three dimensions},
	volume = {125},
	issn = {0035-9009, 1477-870X},
	doi = {10.1002/qj.49712555417},
	language = {en},
	number = {554},
	journal = {Quarterly Journal of the Royal Meteorological Society},
	author = {Gaspari, Gregory and Cohn, Stephen E.},
	month = jan,
	year = {1999},
	pages = {723--757},
}

@article{bishop_gain_2017,
	title = {Gain {Form} of the {Ensemble} {Transform} {Kalman} {Filter} and {Its} {Relevance} to {Satellite} {Data} {Assimilation} with {Model} {Space} {Ensemble} {Covariance} {Localization}},
	volume = {145},
	copyright = {http://www.ametsoc.org/PUBSReuseLicenses},
	issn = {0027-0644, 1520-0493},
	doi = {10.1175/MWR-D-17-0102.1},
	language = {en},
	number = {11},
	journal = {Monthly Weather Review},
	author = {Bishop, Craig H. and Whitaker, Jeffrey S. and Lei, Lili},
	month = nov,
	year = {2017},
	pages = {4575--4592},
}

@article{price_probabilistic_2025,
	title = {Probabilistic weather forecasting with machine learning},
	volume = {637},
	issn = {0028-0836, 1476-4687},
	doi = {10.1038/s41586-024-08252-9},
	language = {en},
	number = {8044},
	journal = {Nature},
	author = {Price, Ilan and Sanchez-Gonzalez, Alvaro and Alet, Ferran and Andersson, Tom R. and El-Kadi, Andrew and Masters, Dominic and Ewalds, Timo and Stott, Jacklynn and Mohamed, Shakir and Battaglia, Peter and Lam, Remi and Willson, Matthew},
	month = jan,
	year = {2025},
	pages = {84--90},
}

@article{sun_all-sky_2025,
	title = {All-sky {AMSU}-{A} radiance data assimilation using the gain-form of {Local} {Ensemble} {Transform} {Kalman} filter within {MPAS}-{JEDI}-2.1.0: implementation, tuning, and evaluation},
	volume = {18},
	copyright = {https://creativecommons.org/licenses/by/4.0/},
	issn = {1991-9603},
	shorttitle = {All-sky {AMSU}-{A} radiance data assimilation using the gain-form of {Local} {Ensemble} {Transform} {Kalman} filter within {MPAS}-{JEDI}-2.1.0},
	doi = {10.5194/gmd-18-8569-2025},
	language = {en},
	number = {22},
	journal = {Geoscientific Model Development},
	author = {Sun, Tao and Guerrette, Jonathan J. and Liu, Zhiquan and Ban, Junmei and Jung, Byoung-Joo and Hernandez Banos, Ivette and Snyder, Chris},
	month = nov,
	year = {2025},
	pages = {8569--8587},
}

@article{slivinski_assimilating_2025,
	title = {Assimilating {Observed} {Surface} {Pressure} {Into} {ML} {Weather} {Prediction} {Models}},
	volume = {52},
	issn = {0094-8276, 1944-8007},
	doi = {10.1029/2024GL114396},
	language = {en},
	number = {6},
	journal = {Geophysical Research Letters},
	author = {Slivinski, L. C. and Whitaker, J. S. and Frolov, S. and Smith, T. A. and Agarwal, N.},
	month = mar,
	year = {2025},
	pages = {e2024GL114396},
}

@article{bano-medina_are_2025,
	title = {Are {AI} weather models learning atmospheric physics? {A} sensitivity analysis of cyclone {Xynthia}},
	volume = {8},
	issn = {2397-3722},
	shorttitle = {Are {AI} weather models learning atmospheric physics?},
	doi = {10.1038/s41612-025-00949-6},
	language = {en},
	number = {1},
	author = {Baño-Medina, Jorge and Sengupta, Agniv and Doyle, James D. and Reynolds, Carolyn A. and Watson-Parris, Duncan and Monache, Luca Delle},
	month = mar,
	year = {2025},
	pages = {92},
}

@misc{pathak_fourcastnet_2022,
	title = {{FourCastNet}: {A} {Global} {Data}-driven {High}-resolution {Weather} {Model} using {Adaptive} {Fourier} {Neural} {Operators}},
	copyright = {arXiv.org perpetual, non-exclusive license},
	shorttitle = {{FourCastNet}},
	doi = {10.48550/ARXIV.2202.11214},
	publisher = {arXiv},
	author = {Pathak, Jaideep and Subramanian, Shashank and Harrington, Peter and Raja, Sanjeev and Chattopadhyay, Ashesh and Mardani, Morteza and Kurth, Thorsten and Hall, David and Li, Zongyi and Azizzadenesheli, Kamyar and Hassanzadeh, Pedram and Kashinath, Karthik and Anandkumar, Animashree},
	year = {2022},
	note = {Version Number: 1},
}

@article{bi_accurate_2023,
	title = {Accurate medium-range global weather forecasting with {3D} neural networks},
	volume = {619},
	copyright = {2023 The Author(s)},
	issn = {1476-4687},
	doi = {10.1038/s41586-023-06185-3},
	language = {en},
	number = {7970},
	journal = {Nature},
	publisher = {Nature Publishing Group},
	author = {Bi, Kaifeng and Xie, Lingxi and Zhang, Hengheng and Chen, Xin and Gu, Xiaotao and Tian, Qi},
	month = jul,
	year = {2023},
	pages = {533--538},
}

@inproceedings{bonev_spherical_2023,
	title = {Spherical {Fourier} {Neural} {Operators}: {Learning} {Stable} {Dynamics} on the {Sphere}},
	issn = {2640-3498},
	shorttitle = {Spherical {Fourier} {Neural} {Operators}},
	url = {https://proceedings.mlr.press/v202/bonev23a.html},
	language = {en},
	urldate = {2026-07-09},
	booktitle = {Proceedings of the 40th {International} {Conference} on {Machine} {Learning}},
	publisher = {PMLR},
	author = {Bonev, Boris and Kurth, Thorsten and Hundt, Christian and Pathak, Jaideep and Baust, Maximilian and Kashinath, Karthik and Anandkumar, Anima},
	month = jul,
	year = {2023},
	pages = {2806--2823},
}

@misc{lang_aifs_2024,
	title = {{AIFS} -- {ECMWF}'s data-driven forecasting system},
	doi = {10.48550/arXiv.2406.01465},
	publisher = {arXiv},
	author = {Lang, Simon and Alexe, Mihai and Chantry, Matthew and Dramsch, Jesper and Pinault, Florian and Raoult, Baudouin and Clare, Mariana C. A. and Lessig, Christian and Maier-Gerber, Michael and Magnusson, Linus and Bouallègue, Zied Ben and Nemesio, Ana Prieto and Dueben, Peter D. and Brown, Andrew and Pappenberger, Florian and Rabier, Florence},
	month = aug,
	year = {2024},
	note = {arXiv:2406.01465 [physics.ao-ph]},
}

@techreport{yessad_full-pos_2019,
	title = {{FULL}-{POS} in the cycle 46t1r1 of {ARPEGE}/{IFS}},
	url = {https://www.cnrm.meteo.fr/gmapdoc/spip.php?article157},
	institution = {Meteo France},
	author = {Yessad, Karim},
	year = {2019},
}

@article{buehner_ensemble_2017,
	chapter = {Monthly Weather Review},
	title = {An {Ensemble} {Kalman} {Filter} for {Numerical} {Weather} {Prediction} {Based} on {Variational} {Data} {Assimilation}: {VarEnKF}},
	volume = {145},
	issn = {1520-0493, 0027-0644},
	shorttitle = {An {Ensemble} {Kalman} {Filter} for {Numerical} {Weather} {Prediction} {Based} on {Variational} {Data} {Assimilation}},
	doi = {10.1175/MWR-D-16-0106.1},
	language = {EN},
	number = {2},
	journal = {Monthly Weather Review},
	publisher = {American Meteorological Society},
	author = {Buehner, Mark and McTaggart-Cowan, Ron and Heilliette, Sylvain},
	month = feb,
	year = {2017},
	pages = {617--635},
}

@article{lorenc_computing_2017,
	title = {Computing an ensemble of variational data assimilations using its mean and perturbations},
	volume = {143},
	copyright = {© 2016 Crown Copyright, Met Office. Quarterly Journal of the Royal Meteorological Society © 2016 Royal Meteorological Society},
	issn = {1477-870X},
	doi = {10.1002/qj.2965},
	language = {en},
	number = {703},
	journal = {Quarterly Journal of the Royal Meteorological Society},
	author = {Lorenc, Andrew C. and Jardak, Mohamed and Payne, Tim and Bowler, Neill E. and Wlasak, Marek A.},
	year = {2017},
	note = {\_eprint: https://rmets.onlinelibrary.wiley.com/doi/pdf/10.1002/qj.2965},
	pages = {798--805},
}

@article{parrish_national_1992,
	chapter = {Monthly Weather Review},
	title = {The {National} {Meteorological} {Center}'s {Spectral} {Statistical}-{Interpolation} {Analysis} {System}},
	volume = {120},
	issn = {1520-0493, 0027-0644},
	doi = {10.1175/1520-0493(1992)120<1747:TNMCSS>2.0.CO;2},
	language = {EN},
	number = {8},
	journal = {Monthly Weather Review},
	publisher = {American Meteorological Society},
	author = {Parrish, David F. and Derber, John C.},
	month = aug,
	year = {1992},
	pages = {1747--1763},
}

@article{tian_evaluating_2026,
	title = {Evaluating {Machine} {Learning} {Weather} {Models} for {Data} {Assimilation}: {Fundamental} {Limitations} in {Tangent} {Linear} and {Adjoint} {Properties}},
	volume = {53},
	copyright = {© 2026. The Author(s).},
	issn = {1944-8007},
	shorttitle = {Evaluating {Machine} {Learning} {Weather} {Models} for {Data} {Assimilation}},
	doi = {10.1029/2025GL119402},
	language = {en},
	number = {2},
	journal = {Geophysical Research Letters},
	author = {Tian, Xiaoxu and Holdaway, Daniel and Kleist, Daryl},
	year = {2026},
	note = {\_eprint: https://agupubs.onlinelibrary.wiley.com/doi/pdf/10.1029/2025GL119402},
	pages = {e2025GL119402},
}

@article{adrian_data_2025,
	chapter = {Artificial Intelligence for the Earth Systems},
	title = {Data {Assimilation} with {Machine} {Learning} {Surrogate} {Models}: {A} {Case} {Study} with {FourCastNet}},
	volume = {4},
	issn = {2769-7525},
	shorttitle = {Data {Assimilation} with {Machine} {Learning} {Surrogate} {Models}},
	doi = {10.1175/AIES-D-24-0050.1},
	language = {EN},
	number = {3},
	journal = {Artificial Intelligence for the Earth Systems},
	publisher = {American Meteorological Society},
	author = {Adrian, Melissa and Sanz-Alonso, Daniel and Willett, Rebecca},
	month = jul,
	year = {2025},
}

@article{kotsuki_ensemble_2025,
	title = {Ensemble data assimilation to diagnose {AI}-based weather prediction models: a case with {ClimaX} version 0.3.1},
	volume = {18},
	issn = {1991-959X},
	shorttitle = {Ensemble data assimilation to diagnose {AI}-based weather prediction models},
	doi = {10.5194/gmd-18-7215-2025},
	language = {English},
	number = {20},
	journal = {Geoscientific Model Development},
	publisher = {Copernicus GmbH},
	author = {Kotsuki, Shunji and Shiraishi, Kenta and Okazaki, Atsushi},
	month = oct,
	year = {2025},
	pages = {7215--7225},
}

@misc{takeshima_bridging_2025,
	title = {Bridging {Artificial} {Intelligence} and {Data} {Assimilation}: {The} {Data}-driven {Ensemble} {Forecasting} {System} {ClimaX}-{LETKF}},
	shorttitle = {Bridging {Artificial} {Intelligence} and {Data} {Assimilation}},
	doi = {10.48550/arXiv.2512.14444},
	publisher = {arXiv},
	author = {Takeshima, Akira and Shiraishi, Kenta and Okazaki, Atsushi and Tsuyuki, Tadashi and Kotsuki, Shunji},
	month = dec,
	year = {2025},
	note = {arXiv:2512.14444 [cs.LG]},
}

@article{huang_ensemble-based_2026,
	title = {Ensemble-{Based} {Assimilation} of {Sounding} {Observations} {With} {AI} {Weather} {Models}},
	volume = {53},
	copyright = {© 2026 The Author(s).},
	issn = {1944-8007},
	doi = {10.1029/2026GL122153},
	language = {en},
	number = {11},
	journal = {Geophysical Research Letters},
	author = {Huang, Huifeng and Lei, Lili and Tan, Zhe-Min},
	year = {2026},
	note = {\_eprint: https://agupubs.onlinelibrary.wiley.com/doi/pdf/10.1029/2026GL122153},
	pages = {e2026GL122153},
}

@misc{xiao_fengwu-4dvar_2024,
	title = {{FengWu}-{4DVar}: {Coupling} the {Data}-driven {Weather} {Forecasting} {Model} with {4D} {Variational} {Assimilation}},
	shorttitle = {{FengWu}-{4DVar}},
	doi = {10.48550/arXiv.2312.12455},
	publisher = {arXiv},
	author = {Xiao, Yi and Bai, Lei and Xue, Wei and Chen, Kang and Han, Tao and Ouyang, Wanli},
	month = may,
	year = {2024},
	note = {arXiv:2312.12455 [physics.ao-ph]},
}

@misc{wang_mapcast_2026,
	title = {{MAPCast}: {A} {Convection} {Allowing} {MPAS} {Emulator} for {Ensemble}-based {Background} {Error} {Covariance} {Estimation} {Toward} {Multi}-{Scale} {Data} {Assimilation}},
	shorttitle = {{MAPCast}},
	doi = {10.48550/arXiv.2607.21917},
	publisher = {arXiv},
	author = {Wang, Yongming and Wang, Xuguang},
	month = jul,
	year = {2026},
	note = {arXiv:2607.21917 [physics.ao-ph]},
}

@article{xu_fuxi-da_2025,
	title = {{FuXi}-{DA}: a generalized deep learning data assimilation framework for assimilating satellite observations},
	volume = {8},
	copyright = {2025 The Author(s)},
	issn = {2397-3722},
	shorttitle = {{FuXi}-{DA}},
	doi = {10.1038/s41612-025-01039-3},
	language = {en},
	number = {1},
	journal = {npj Climate and Atmospheric Science},
	publisher = {Nature Publishing Group},
	author = {Xu, Xiaoze and Sun, Xiuyu and Han, Wei and Zhong, Xiaohui and Chen, Lei and Gao, Zhiqiu and Li, Hao},
	month = apr,
	year = {2025},
	pages = {156},
}

@article{sun_data--forecast_2025,
	title = {A data-to-forecast machine learning system for global weather},
	volume = {16},
	issn = {2041-1723},
	doi = {10.1038/s41467-025-62024-1},
	language = {en},
	number = {1},
	journal = {Nature Communications},
	author = {Sun, Xiuyu and Zhong, Xiaohui and Xu, Xiaoze and Huang, Yuanqing and Li, Hao and Neelin, J. David and Chen, Deliang and Feng, Jie and Han, Wei and Wu, Libo and Qi, Yuan},
	month = jul,
	year = {2025},
	pages = {6658},
}

@article{bonavita_limitations_2024,
	title = {On {Some} {Limitations} of {Current} {Machine} {Learning} {Weather} {Prediction} {Models}},
	volume = {51},
	copyright = {© 2024 ECMWF.},
	issn = {1944-8007},
	doi = {10.1029/2023GL107377},
	language = {en},
	number = {12},
	journal = {Geophysical Research Letters},
	author = {Bonavita, Massimo},
	year = {2024},
	pages = {e2023GL107377},
}

@misc{subich_fixing_2025,
	title = {Fixing the {Double} {Penalty} in {Data}-{Driven} {Weather} {Forecasting} {Through} a {Modified} {Spherical} {Harmonic} {Loss} {Function}},
	doi = {10.48550/arXiv.2501.19374},
	publisher = {arXiv},
	author = {Subich, Christopher and Husain, Syed Zahid and Separovic, Leo and Yang, Jing},
	month = may,
	year = {2025},
	note = {arXiv:2501.19374 [cs.LG]},
}

@misc{goecke_aicon_2026,
	title = {{AICON}: {An} operational global machine learning weather forecasting model},
	shorttitle = {{AICON}},
	doi = {10.48550/arXiv.2608.24651},
	publisher = {arXiv},
	author = {Goecke, Tobias and Jacob, Marek and Prill, Florian and Denhard, Michael and Fundel, Felix and Keller, Jan and Potthast, Roland and Reich, Hendrik and Seegebrecht, Britta and Ulbrich, Sven and Valmassoi, Arianna and Wahl, Sabrina},
	month = aug,
	year = {2026},
	note = {arXiv:2608.24651 [physics.ao-ph]},
}

@article{courtier_strategy_1994,
	title = {A strategy for operational implementation of {4D}-{Var}, using an incremental approach},
	volume = {120},
	copyright = {Copyright © 1994 Royal Meteorological Society},
	issn = {1477-870X},
	doi = {10.1002/qj.49712051912},
	language = {en},
	number = {519},
	journal = {Quarterly Journal of the Royal Meteorological Society},
	author = {Courtier, P. and Thépaut, J.-N. and Hollingsworth, A.},
	year = {1994},
	pages = {1367--1387},
}

@article{derber_global_1989,
	chapter = {Journal of Physical Oceanography},
	title = {A {Global} {Oceanic} {Data} {Assimilation} {System}},
	volume = {19},
	issn = {0022-3670, 1520-0485},
	doi = {10.1175/1520-0485(1989)019<1333:AGODAS>2.0.CO;2},
	language = {EN},
	number = {9},
	journal = {Journal of Physical Oceanography},
	publisher = {American Meteorological Society},
	author = {Derber, John and Rosati, Anthony},
	month = sep,
	year = {1989},
	pages = {1333--1347},
}

@article{whitaker_evaluating_2012,
	chapter = {Monthly Weather Review},
	title = {Evaluating {Methods} to {Account} for {System} {Errors} in {Ensemble} {Data} {Assimilation}},
	volume = {140},
	issn = {1520-0493, 0027-0644},
	doi = {10.1175/MWR-D-11-00276.1},
	language = {EN},
	number = {9},
	journal = {Monthly Weather Review},
	publisher = {American Meteorological Society},
	author = {Whitaker, Jeffrey S. and Hamill, Thomas M.},
	month = sep,
	year = {2012},
	pages = {3078--3089},
}

@article{zhang_impacts_2004,
	chapter = {Monthly Weather Review},
	title = {Impacts of {Initial} {Estimate} and {Observation} {Availability} on {Convective}-{Scale} {Data} {Assimilation} with an {Ensemble} {Kalman} {Filter}},
	volume = {132},
	issn = {1520-0493, 0027-0644},
	doi = {10.1175/1520-0493(2004)132<1238:IOIEAO>2.0.CO;2},
	language = {EN},
	number = {5},
	journal = {Monthly Weather Review},
	publisher = {American Meteorological Society},
	author = {Zhang, F. and Snyder, Chris and Sun, Juanzhen},
	month = may,
	year = {2004},
	pages = {1238--1253},
}

%
%
%
%
%

\end{document}


\maketitle
{\renewcommand{\thefootnote}{}%
 \footnotetext{Corresponding author: Simon Toedtli, \url{stoedtli@ucar.edu}}}

\noindent\textbf{Contents of this file}
\begin{enumerate}
\item Text S1 to S5
\item Figures S1 to S3
\item Table S1
\end{enumerate}

\bigskip
\noindent\textbf{Introduction}

\begin{figure}
    \noindent\includegraphics[width=0.95\linewidth]{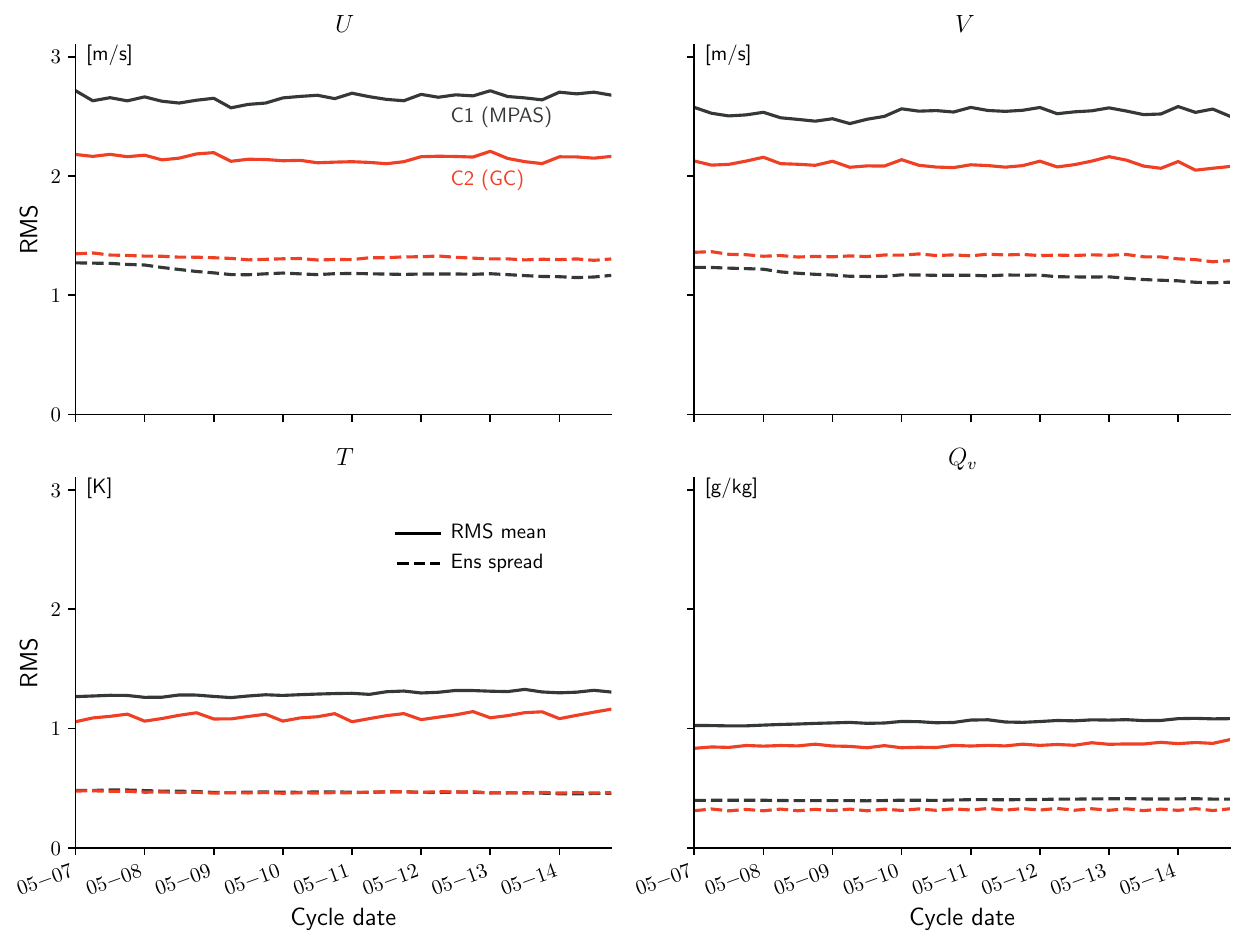}
    \caption{Time series of RMS-spread statistics for the EnKF background ensemble over the last week of cycling. Solid lines show the RMSE of the ensemble mean against the GFS analysis, dashed lines represent the ensemble spread. Black: configuration C1 (MPAS), red: configuration C2 (GC). Results are averaged horizontally and vertically over the levels not directly affected by spurious correlations near the GC model top (up to model level 34, indicated by the dotted magenta line in figs. 2 and 3).}
    \label{fig:ensSpreadTimeseries}
\end{figure}

\begin{figure}
    \noindent\includegraphics[width=\linewidth]{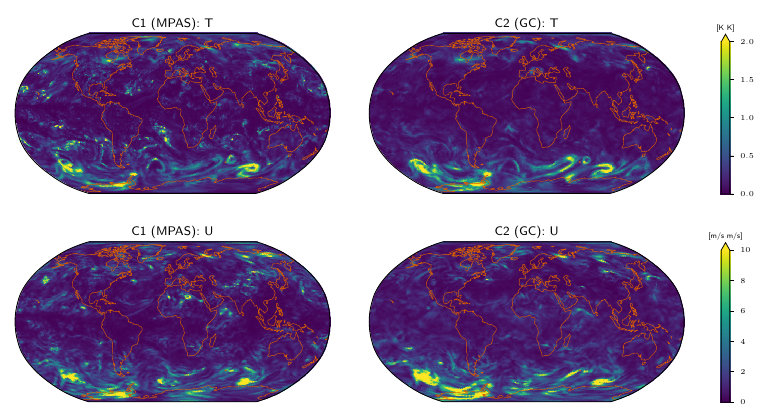}
    \caption{Ensemble spread at model level 17 at the end of the cycling period (2018-05-15T00). Top row: temperature, bottom row: zonal wind. Left column: configuration C1 (MPAS), right column: C2 (GC).}
    \label{fig:ensSpreadSpatial}
\end{figure}

\begin{figure}
    \noindent\includegraphics[width=0.95\linewidth]{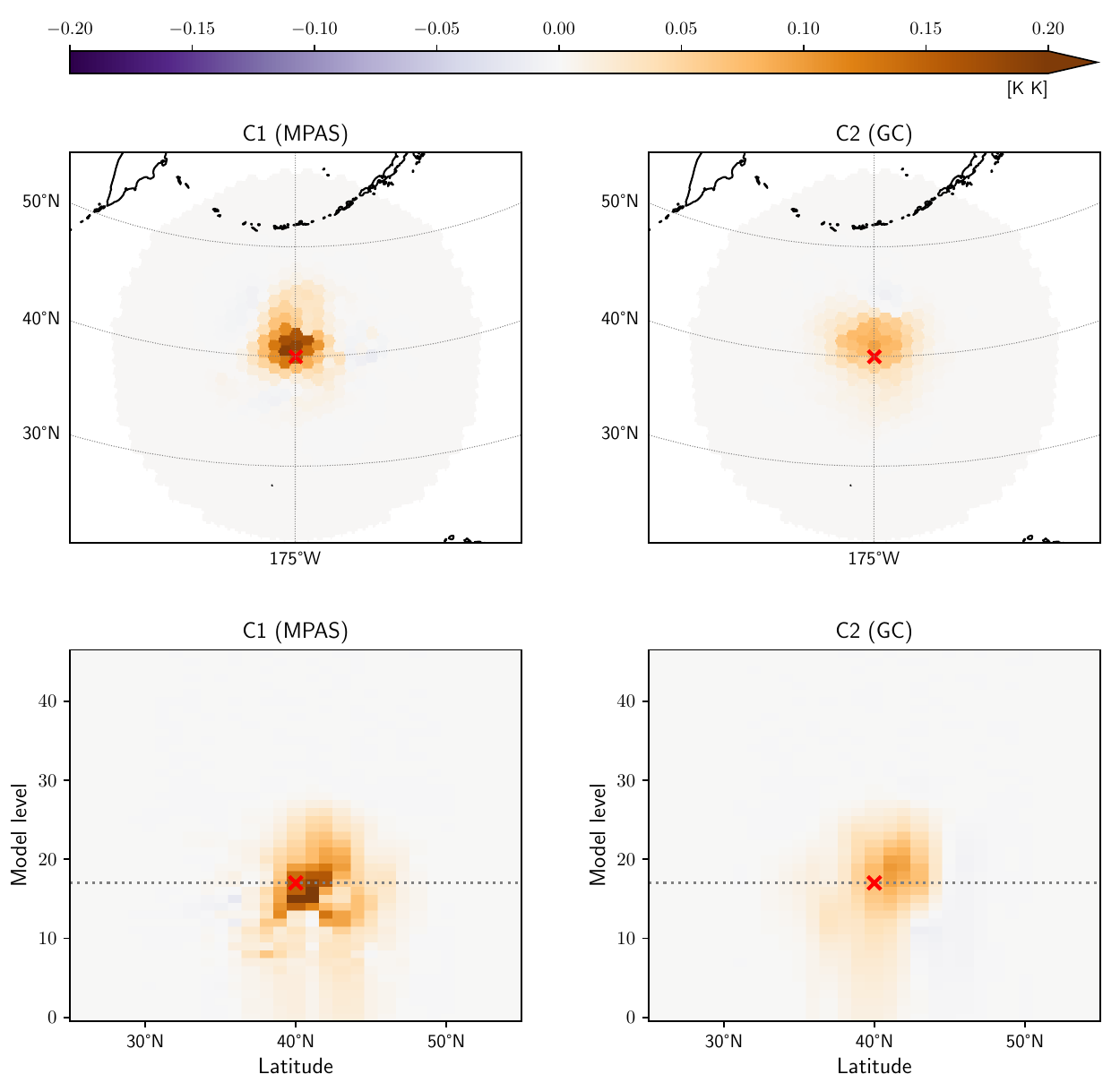}
    \caption{Single-observation experiment at the end of the cycling period (2018-05-15T00): autocovariance of the temperature field, $\mathrm{cov}(T(\bm{r}), T(\bm{r}_M))$, where $\bm{r}_M$ denotes the measurement location. The measurement location is at $40\degree$N, $175\degree$W, $702$ hPa and indicated by the $\times$ marker. Left column: C1 (MPAS); right column: C2 (GC). Top row: horizontal cross section at the measurement height; bottom row: vertical cross section along $175\degree$W.}
    \label{fig:covTT}
\end{figure}

This supplementary material provides additional details and figures to support the conclusions in the main manuscript.
Texts S1–S5 describe, respectively, the concept of cycling in DA systems, the DA components, ensemble inflation, the interpolation between the GC and MPAS meshes, and the assimilated observations.
The assimilated microwave sensors and channels are listed in \cref{table:assimRadiances}, while \cref{fig:ensSpreadTimeseries,fig:ensSpreadSpatial,fig:covTT} provide additional figures of ensemble spread and covariance structure.
\clearpage

\noindent\textbf{Text S1. Cycling in data assimilation workflows}

Data assimilation for weather prediction typically employs ``cycling’’: DA is performed at a regular cadence, for example every six hours in this study.
The inputs to the DA scheme are the most recent observations and a prior forecast from the previous cycle (background), and the output is a new initial condition (analysis).
The analysis then initializes the forecast that provides the background for the next cycle.

\bigskip
\noindent\textbf{Text S2. EnVar and GETKF Components}

The configuration of the DA components closely follows previous MPAS-JEDI studies:
the EnVar component adopts the clear-sky configuration of \citet{liu_data_2022}, while the GETKF component follows the clear-sky configuration of \citet{sun_all-sky_2025}.
An in-depth discussion can be found in those references, the following paragraphs highlight a few aspects that are most relevant to this study.

The EnVar component uses a purely ensemble-based background error covariance model.
This is a deliberate choice, as it provides the most stringent test for the ensemble forecast models that evolve these covariances.
The ensemble-based background error covariance is localized with a Gaspari–Cohn function \citep{gaspari_construction_1999}, using a horizontal and vertical full-width of $1200$ km and $6$ km, respectively, beyond which correlations are set to zero.
The 3D-EnVar cost function is solved in incremental form \citep{courtier_strategy_1994} with two outer loops and $60$ inner loops, via the Derber–Rosati Preconditioned Conjugate Gradient (DRPCG) algorithm \citep{derber_global_1989}.

For the GETKF component, horizontal localization is performed in observation space using the R-localization approach \citep{hunt_efficient_2007} with a localization length scale of $1200$ km.
Vertical localization, in contrast, is performed in model space by modulating the original ensemble members with the leading eigenvectors of a localization matrix \citep{bishop_gain_2017}.
Here, a Gaspari–Cohn localization matrix with a 6 km full-width is used and the leading $9$ eigenvectors are retained, which capture approximately 95\% of the variance of the localization matrix.

\bigskip
\noindent\textbf{Text S3. Ensemble Inflation}

Following \citet{sun_all-sky_2025}, two posterior ensemble inflation strategies are employed in the present study:
relaxation to prior perturbations (RTPP) and relaxation to prior spread (RTPS).

The RTPP scheme \citep{zhang_impacts_2004} replaces the analysis perturbations $\bm{\delta x}_i^a$ of member $i$ with a linear combination of the analysis and background perturbations, $\bm{\delta x}_i^b$
\begin{equation}
    \bm{\delta x}_i^a \leftarrow \alpha_{rtpp} \, \bm{\delta x}_i^b + \left( 1 - \alpha_{rtpp} \right) \bm{\delta x}_i^a
\end{equation}
where $\alpha_{rtpp}$ is a scalar relaxation parameter.

The RTPS scheme \citep{whitaker_evaluating_2012}, in contrast, relaxes the spread of the analysis ensemble towards the spread of the background ensemble.
For each grid cell and variable, jointly indexed by $j$, the spread change ratio $s_j$ is calculated as
\begin{equation}
    s_j = \frac{\sigma^b_j - \sigma^a_j}{\sigma^a_j}
\end{equation}
where $\sigma^a$ and $\sigma^b$ are the ensemble standard deviations of the analysis and background, respectively.
The analysis perturbations are then updated as
\begin{equation}
    \bm{\delta x}^a_i \leftarrow \bm{\delta x}^a_i \circ (\alpha_{rtps} \, \bm{s} + \bm{1})
\end{equation}
where $\alpha_{rtps}$ is a relaxation parameter and $\circ$ denotes the Schur product.

Both relaxation parameters are scalars and take on values between $\alpha = 0$ (no inflation) and 1 (full replacement with the background state).
Their optimal values must be determined empirically, and separate tuning experiments were performed for configurations C1 and C2.
Interestingly, the optimal parameters turned out to be identical for both configurations: $\alpha_{rtpp} = 0.5$ and $\alpha_{rtps} = 0.75$.
These settings were used for all experiments shown in the manuscript.

Initial sensitivity tests showed that the filter remains stable over a wide range of parameter settings.
However, it is difficult to compare the performance of configurations C1 and C2 if their underlying ensemble spread differs or evolves differently over time.
For the chosen inflation parameters, the ensemble spread of the two systems is comparable and remains approximately constant over time (see \cref{fig:ensSpreadTimeseries}), with one exception:
during the first two days of cycling, the spread of configuration C2 initially decreases before recovering to the level shown in \cref{fig:ensSpreadTimeseries}.
We have not investigated this initial decrease, but note that similar behavior appears to be common in MLWP-based ensemble systems \citep[see e.g. Figs. 3 and 5 in][]{takeshima_bridging_2025}.

\bigskip
\noindent\textbf{Text S4. Interpolation Between GraphCast and MPAS Mesh}

As mentioned in section 2.2 of the manuscript, all data assimilation steps are performed in MPAS-JEDI on the native MPAS mesh, regardless of the choice of ensemble forecast model.
In configuration C2, this requires interpolating between the MPAS and GC meshes before and after each ensemble forecast.

The mapping from the MPAS to the GC mesh requires interpolation from MPAS's hybrid, terrain-following vertical coordinate to GC's constant pressure levels, including extrapolation below orography in areas of high terrain.
This mapping follows ERA5 conventions \citep[see e.g.][]{yessad_full-pos_2019}, to ensure that the interpolated analyses are consistent with the ERA5 data model used to train GraphCast.

The reverse interpolation, from the GC to the MPAS mesh, uses MPAS's own initialization system.
The interpolation method within the GC domain and extrapolation method above it are variable-specific.
Because GC lacks several fields required by MPAS (skin temperature, sea surface temperature, water-equivalent snow depth, sea ice, soil moisture, and soil temperature), these are taken from the GFS analysis valid at the cycle time and combined with the GC output prior to interpolation.
Similarly, GC's static land-sea mask and terrain height are replaced with those from the GFS analysis, to avoid artifacts due to mismatches in model representation.
In principle, these missing fields could instead be taken from the deterministic EnVar component, which is also evolved with MPAS in both configurations and therefore includes all of these variables.
Once all fields have been interpolated to the MPAS mesh, a single short MPAS time step ($\mathrm{dt} = 1$ min) is performed to compute the diagnostic variables required to evaluate the observation operators in the DA step.
This time step introduces a time drift of $4$ minutes per day, which could become problematic for GC-based configurations run over much longer periods.
However, no apparent detrimental effect from this drift was observed over the month-long experiment presented in this manuscript.

Initial sensitivity experiments explored several model-top arrangements, including a standard $30$ km MPAS model top and a lower $19.6$ km MPAS model top.
Negative interpolation effects were smallest with a $22.6$ km model top, which is just above the GC model top.
This configuration is therefore used throughout the study.

\bigskip
\noindent\textbf{Text S5. Assimilated observations}

The EnKF and EnVar components assimilate the same observations, which represent a substantial subset of the global observation network.
These include temperature, specific humidity, and zonal and meridional wind components from aircraft and radiosondes; surface pressure; zonal and meridional wind components from satellite-derived atmospheric motion vectors; Global Navigation Satellite System Radio Occultation (GNSS-RO) bending angle profiles; and selected clear-sky channels from the Advanced Microwave Sounding Unit-A (AMSU-A) and Microwave Humidity Sounder (MHS).
The assimilated microwave radiance channels are specified in \cref{table:assimRadiances}.
Note that AMSU-A channel $9$, which has been assimilated in previous MPAS-JEDI studies, is omitted here because its weight function peaks close to the GC model top and retains nonzero support well above it.

\bibliographystyle{agufull08}
\bibliography{bibliography}

\clearpage

\begin{table}
\caption{Assimilated AMSU-A and MHS clear-sky channels}
\centering
\setlength{\tabcolsep}{12pt}
\begin{tabular}{llc}
\toprule
Satellite  & Sensor & Clear-sky channel  \\
\midrule
Aqua     & AMSU-A & 8 \\
\midrule
NOAA-15  & AMSU-A & 5, 7, 8 \\
\midrule
\multirow{2}{*}{NOAA-18} & AMSU-A & 5, 6, 7, 8 \\
         & MHS    & 3, 4, 5 \\
\midrule
\multirow{2}{*}{NOAA-19} & AMSU-A & 5, 6, 7 \\
         & MHS    & 3, 4, 5 \\
\midrule
\multirow{2}{*}{Metop-A} & AMSU-A & 5, 6 \\
         & MHS    & 3, 4, 5 \\
\midrule
\multirow{2}{*}{Metop-B} & AMSU-A & 8 \\
         & MHS    & 3, 4, 5 \\
\bottomrule
\end{tabular}
\label{table:assimRadiances}
\end{table}

